\documentstyle[aps,preprint,epsf]{revtex}
\def\inseps#1#2{\def\epsfsize##1##2{#2##1} \centerline{\epsfbox{#1}}}

\newcommand{\spmine}{1.20}
\newcommand{\myspace}{\edef\baselinestretch{\spmine}\Large\normalsize}
\begin{document}
\myspace

%\draft

\title{Exact zero-point energy shift in the $e\otimes (n~E)$,
$t\otimes (n~H)$ many modes dynamic Jahn-Teller systems at strong coupling}

\author{Nicola Manini\thanks{Electronic address: manini@esrf.fr}}
\address{E.S.R.F., B.P. 220, F 38043 Grenoble Cedex, France}
\author{Erio Tosatti\thanks{Electronic address: tosatti@sissa.it} }
\address{
International School for Advanced Studies (SISSA), via Beirut 2-4, I-34014
Trieste, Italy\\ Istituto Nazionale Fisica della Materia, Unit\`a Trieste
SISS and\\ International Centre for Theoretical Physics (ICTP),
P.O. BOX 586, I-34014 Trieste, Italy}

\maketitle

\begin{abstract}
We find the exact semiclassical (strong coupling) zero-point energy shifts
applicable to the $e\otimes (n\ E)$ and $t\otimes (n\ H)$ dynamic
Jahn-Teller problems, for an arbitrary number $n$ of discrete vibrational
modes simultaneously coupled to one single electronic level. We also obtain
an analytical formula for the frequency of the resulting normal modes,
which has an attractive and apparently general Slater-Koster form. The
limits of validity of this approach are assessed by comparison with
O'Brien's previous effective-mode approach, and with accurate numerical
diagonalizations.
Numerical values obtained for $t\otimes (n\ H)$ with $n =8$ and coupling
constants appropriate to C$_{60}^-$ are used for this purpose, and are
discussed in the context of fullerene.

\end{abstract}

%31.15.Gy Semiclassical methods of Electronic structure of atoms, molecules
%	  and their ions: theory
%33.20.Wr Vibronic, rovibronic, and rotation-electron-spin interactions
%36.40.Mr Spectroscopy and geometrical structure of clusters
%61.46.+w Clusters, nanoparticles, and nanocrystalline materials
%61.48.+c Fullerenes and fullerene-related materials
%63.20.Kr Phonon-electron  and phonon-phonon interactions
%71.38.+i Polarons and electron-phonon interactions (see also 63.20.K Phonon-
%         electron interactions in lattices)

\section{Introduction}

The dynamical Jahn-Teller (DJT) effect, where one degenerate electronic
level is nontrivially coupled to vibrational modes, represents perhaps the
simplest type of problem where the Born-Oppenheimer approximation is
invalid in principle, and a quantum mechanical treatment of the full
electron-ion system is essential. Even for the simplest DJT case,
true closed-form solutions have only been
obtained for a restricted set of coupling values\cite{Szopa}.
In spite of that, the DJT physics of a {\em single} vibrational mode
coupled to the electronic state
is quite well understood\cite{Englman} for all known symmetries, through
analytic expansions valid in the opposite limits of strong and weak
coupling, and otherwise through numerical diagonalization for intermediate
values of the coupling. In a large
number of cases of practical interest there are however {\em many}
vibrational modes, coupled simultaneously to the same degenerate
electronic state.  This many-modes case has been extensively studied
theoretically\cite{slonczewski,ob72,Bersuker}.  In particular, a first
group of studies considers in detail the case of an impurity-related
localized degenerate level interacting with the continuum of phonons in a
crystal\cite{slonczewski,Bersuker}, while a second class of studies
concentrates on the case of a finite, discrete set of vibrational
modes\cite{ob72,ob80,ob83}.  In the present paper we revisit the second
type of problem, having in mind in particular the case of an isolated
molecule/cluster/ion with a finite (although possibly large) number $n$ of
vibrational modes participating.

To begin with, we recall the two special limits in which the
many-modes problem is
trivially solved\cite{Bersuker}.  First, the case
of all modes having the same symmetry
and frequency, is simply reduced, by means of a
rotation in the vibrational space, to the equivalent problem of a single
mode coupled with the total coupling intensity, plus $n-1$ uncoupled
modes.  Secondly, in the weak-coupling limit, degenerate
perturbation theory applies, and contributions from different modes
linearly superpose.  Intermode interactions appear only at
fourth and higher order in the couplings.

The more realistic case of many modes with the same symmetry and different
frequencies and couplings was apparently first addressed systematically by
O'Brien\cite{ob72} in the other customary limit, i.e., the strong-coupling
semiclassic expansion.  Her approach relies on the concept of replacing the
$n$ real modes with a single effective mode, perturbed by weak residual
off-diagonal contributions.  In this way one obtains an approximation which
has been very useful, notably\cite{ob83} in the calculation of spectral
shapes and reduction factors.  Nonetheless, as we will show below, there is
no well-defined limiting case where that approximation become exact in a
controlled manner, except for the equal-frequency case.

In this work we demonstrate an alternative approach, which is exact in the
semiclassical (strong e-v coupling) limit, and applicable from strong to
intermediate couplings.  Rather than working in general, we address
directly two specific and physically important cases of many-modes DJT,
namely the linear $e\otimes (n E)$ and $t\otimes (n\ H)$.  Our method
relies on a detailed analysis of the Born-Oppenheimer (BO) potential
surface around its minimum, and of the normal frequencies of the classical
small oscillations of the e-v coupled system.  These frequencies, in turn,
determine residual zero-point quantum correction to the classical
motion on the BO surface.
Comparison with the effective-mode
approach\cite{ob72} reveals that the latter, not taking into account these
corrections, contains a systematic deviation in the strong-coupling
limit, which we discuss in detail in Sec.~\ref{numerical:sect}. Inclusion
of these corrections generates the leading term of the exact
strong-coupling expansion of the JT energy gain as a function of inverse
coupling (Eq.~(\ref{Emmodes:eqn})). As a very useful byproduct, we
also obtain a detailed description of the
low-energy vibronic spectrum, which is again accurate in the strong
coupling limit.
Finally, in order to provide a numerical application, we deliberately
choose in Sec.~\ref{txH:sect} a borderline case of intermediate
coupling, namely the negative ion of C$_{60}$. That is a $t\otimes (n\ H)$
DJT problem, with $n = 8$, for which both exact and effective mode
alternative calculations are feasible.  Due to intermediate coupling,
our approximation will not, of course, be particularly accurate.
Nonetheless, being able to judge the sign and size of the deviation is
especially useful in assessing errors, particularly in a case
like this, which stretches a little beyond
the reasonable borders of validity.

\section{The $e\otimes (n E)$ system}
\label{exE:sect}

We consider first the $e\otimes E$ linear JT Hamiltonian, a basic
textbook\cite{Englman,Bersuker,Abragam} example of DJT, as well as a
relevant model in many molecules and crystals.  For convenience, we shall
adopt the notations of Ref.~\cite{ob72}. The Hamiltonian operator for the
$n$-modes problem is
\begin{equation}
\label{hamiltonian:eqn}
H = \frac 12 \sum_i \omega_i \left(\vec p_i^2 + \vec q_i^2 \right) +
	\sum_\sigma
\left(c_{x\sigma}^\dagger,c_{y\sigma}^\dagger \right) \sum_i k_i \omega_i
\pmatrix{
q_{i1}	&q_{i2} \cr
q_{i2}	&-q_{i1}
}
\pmatrix{c_{x\sigma}\cr c_{y\sigma} } \ ,
\end{equation}
where $\omega_i$ is the frequency and $k_i$ is the dimensionless coupling
strength of the twofold-degenerate mode $i$.  $\vec q_i$ is a vector
notation for the two normal coordinates $q_{i1}$ and $q_{i2}$ of mode $i$,
and $\vec p_i$ are the corresponding conjugate momenta.  $x$ and $y$ label
the two degenerate electronic states.  The sums should be understood
as $\sum_{i=1}^n$.  Note that we use the
second-quantized notation for the fermions, and the coordinate description
for the vibrational degrees of freedom. We also set $\hbar=1$, thus making
no distinctions between (angular) frequencies and energies.

As a first step, we treat the $\vec q_i$ as classical coordinates,
and study the (lowest) Born-Oppenheimer potential surface:
\begin{equation}
\label{BOpotential:eqn}
V\left(\vec q_i\right), =
\frac 12 \sum_i \omega_i (\vec q_i)^2 +
	\min_\psi \left<\psi\right|
	\sum_i k_i \omega_i \left(q_{i1} (c_x^\dagger c_x - c_y^\dagger c_y)
	+ q_{i2} (c_x^\dagger c_y + c_y^\dagger c_x) \right) \left|\psi\right>
\end{equation}
The minimum over the fermionic degree of freedom $\left|\psi\right>$ is the
lowest eigenvalue of the 2$\times$2 electronic problem representing one
fermion in the degenerate level, which is
$-\left(\sum_i k_i^2 \omega_i^2 {\vec
q_i^2} + 2 \sum_{i<j} k_i k_j \omega_i \omega_j {\vec q_i} \cdot {\vec q_j}
\right)^{1/2}$.  The lowest BO surface is conveniently rewritten in polar
coordinates ${\vec q_i}= (q_i \cos{\theta_i}, q_i \sin{\theta_i})$ as
\begin{equation}
\label{BOpot:eqn}
V\left(q_i,\theta_i\right) = \frac 12 \sum_i \omega_i q_i^2 -
	\left\{\sum_i k_i^2 \omega_i^2 q_i^2 +
  2 \sum_{i<j} k_i k_j \omega_i \omega_j q_i q_j \cos(\theta_i-\theta_j)
	\right\}^{1/2}
\end{equation}
In this form, the minima of $V$, corresponding to the classical
stable equilibrium configurations of the system, are straightforwardly
discussed. In
order to minimize $V$, the argument of the square root should be maximum,
which is obtained when all $\theta_i-\theta_j$ simultaneously vanish.
In this case, for all cosines equal to one, the square root can be
explicitly executed, to
obtain separate dependences on the different $q_i$'s. The
minimum of $V$ is therefore obtained for
\begin{eqnarray}
\label{minima:eqn}
q_i^{\rm min}&=&k_i\nonumber \\
\theta_i^{\rm min}&=&\theta,
\end{eqnarray}
where $\theta$ is arbitrary. As form (\ref{BOpot:eqn}) explicitly
shows, the potential energy is independent of the
common rotation angle $\theta=\sum_i \theta_i /n$.  This implies in
particular that the configurations of minimum potential energy constitute a
continuous manifold, parametrized by $\theta$, with the topology of a
circle, exactly the same as in the one-mode problem.  This result was
to be expected also on the basis of more abstract considerations
\cite{Ceulemans}, and we shall return to it below, when analyzing the
small oscillations around the minimum.

The value of the potential at the minimum is the classical JT
energy gain
\begin{equation}
\label{Eclass:eqn}
E_{\rm class} =
- \frac 12 \sum_i k_i^2 \omega_i =
- \frac 12 k_{\rm eff}^2 \omega_{\rm eff} \ ,
\end{equation}
where we have introduced, following Ref.~\cite{ob72}, $k_{\rm eff}^2=\sum_i
k_i^2$ and $\omega_{\rm eff}=\sum_i k_i^2 \omega_i/k_{\rm eff}^2 $.

This was just a re-derivation of well-known results.  The next step is to
compute the quantum corrections to this classical result.  The quantum
corrections add to $E_{\rm class}$ in the form of a strong-coupling
expansion in negative powers of $k_{\rm eff}$.  The leading term, of order
zero, is a shift independent of $k_{\rm eff}$, due to the change of
zero-point energy of the system following the JT coupling.

This is most easily illustrated in the case of just one mode.  The
ground-state energy of Eq.\ (\ref{hamiltonian:eqn}), $E(k_1=0)$ (no JT
coupling), is just the zero-point energy of the (two-fold degenerate) mode,
amounting to $2\cdot \frac 12 \omega_1$.  On the other hand, when the
coupling is very strong, the dynamics factorizes into a radial harmonic
mode plus a free pseudo-rotation.  The harmonic zero-point energy
is now only $\frac 12\omega_1$.  For a single mode, the total JT energy gain
is therefore, as is well known,
\begin{equation}
\label{E1mode:eqn}
E_{\rm JT} - E(k_1=0) =
  E_{\rm class} - \frac 12 \omega_1
+\frac 12 \omega_1 k_1^{-2} j^2 +O\left(k_1^{-4}\right) \ ,
\end{equation}
where $E_{\rm class}=- \frac 12 k_1^2 \omega_1$ term represents the
lowering of the BO potential minimum, $- \frac 12 \omega_1$ representing the
zero-point energy gain, and the $j(=\pm \frac 12$ for the ground state)
term is the residual zero-point energy associated with
quantization of the (pseudo-rotational) $\theta$  motion.  In the rest
of this section we wish to generalize the second
and third term in Eq.\ (\ref{E1mode:eqn}) to the many-modes case.

\subsection{The semiclassical expansion}
\label{results:sect}

We now extend this kind of semiclassical expansion to the many-modes case.
It is clear that the zero-point energy for
zero coupling $E(k_i=0) = 2\times \frac 12\sum_i \omega_i$.
To determine the zero-point
energy for the motion around the many-modes potential minimum in the finite
coupling case, we expand the BO potential to second order about the minimum
in the $\vec{x} \equiv (q_1,q_2,..., \theta_1,\theta_2,...)$ coordinates,
and compute the classical normal modes of vibration:
\begin{equation}
V\left(\vec{x} \right)=
E_{\rm class} + \frac 12 \sum_{\mu,\nu}^{2 n}
	(x_\mu-x_\mu^{\rm min})
	\left.   \frac{\partial^2 V}{\partial x_\mu \partial x_\nu}
	\right|_{\vec{x}^{\rm min}}
	(x_\nu-x_\nu^{\rm min}) +O\left((\vec{x} -\vec{x}^{\rm min})^3 \right)
\label{harpoteq:eqn}
\end{equation}
where the Hessian  matrix of the derivatives is
\begin{equation}
{\bf V}_{\mu \nu} =
\left.   \frac{\partial^2 V}{\partial x_\mu \partial x_\nu}
\right|_{\vec{x}^{\rm min}} =
\left(
\begin{array}{c|c}
	\begin{array}{ccc}
	\omega_1& &\\
	&\omega_2 &\\
	& &\ddots
	\end{array}
& 0 \\
\hline
0 &
	\begin{array}{ccc}
	k_1^2 \omega_1 (1- \Xi k_1^2 \omega_1) &
	- \Xi k_1^2 \omega_1 k_2^2 \omega_2 &...\\
	- \Xi k_1^2 \omega_1 k_2^2 \omega_2&
	k_2^2 \omega_2 (1- \Xi k_2^2 \omega_2) &
	...\\
	...&...&\ddots
	\end{array}
\end{array}
\right)\ ,
\label{hessian:eqn}
\end{equation}
and $\Xi=\left(\sum_i k_i^2 \omega_i\right)^{-1}$.

The kinetic energy for these coordinates is
\begin{equation}
E_{\rm kin}\left(\vec{x}, \dot{\vec{x}} \right)= \frac 12
\sum_{\mu,\nu}^{2 n}
\dot{\vec{x}}_\mu T_{\mu\nu}\left(\vec{x}\right) \dot{\vec{x}}_\nu
\label{kineq:eqn}
\end{equation}
where
\begin{equation}
{\bf T}\left(\vec{x}\right)_{\mu\nu} =
\left(
\begin{array}{c|c}
	\begin{array}{ccc}
	\omega_1^{-1}& &\\
	&\omega_2^{-1} &\\
	& &\ddots
	\end{array}
& 0 \\
\hline
0 &
	\begin{array}{ccc}
	\omega_1^{-1}q_1^2& &\\
	&\omega_2^{-1}q_2^2 &\\
	& &\ddots
	\end{array}
\end{array}
\right)\ .
\label{kinmat:eqn}
\end{equation}
The normal frequencies $\bar\omega^2$ and vibronic modes are obtained as
eigenvalues and eigenvectors of the matrix

\begin{equation}
{\bf T}\left(\vec{x}^{\rm min}\right)^{-1} \cdot {\bf V} =
\left(
\begin{array}{c|c}
	\begin{array}{ccc}
	\omega_1^2& &\\
	&\omega_2^2 &\\
	& &\ddots
	\end{array}
& 0 \\
\hline
0 &
	\begin{array}{ccc}
	\omega_1^2 - \Xi k_1^2 \omega_1^3 &
	- \Xi k_2^2 \omega_2 \omega_1^2 &...\\
	- \Xi k_1^2 \omega_1 \omega_2^2 &
	\omega_2^2 - \Xi k_2^2 \omega_2^3 &
	...\\
	...&...&\ddots
	\end{array}
\end{array}
\right)
\label{dinmat:eqn}
\end{equation}
This matrix is non Hermitian, but has real eigenvalues.
The block-diagonal form shows that the small oscillations of the radial and
angular variables are uncoupled. Each {\it radial}
coordinate $q_i$ corresponds directly to one radial mode of the same
frequency $\omega_i$ as the uncoupled modes.
On the contrary, the dynamics of the {\it angular} variables are intercoupled
through the off-diagonal elements in Eq.\ (\ref{dinmat:eqn}).  The secular
equation for the angular eigenvector of components $x_i$ and eigenfrequency
$\nu^2$ is
\begin{equation}
\nu^2 x_i = \omega_i^2 x_i - \Xi \omega_i^2 \sum_j k_j^2\omega_j  x_j
\label{secular:eqn}
\end{equation}
with the solution
\begin{equation}
x_i = {\rm const} \cdot  \frac {\omega_i^2}{\omega_i^2 -\nu^2}
\label{eigenvector:eqn}
\end{equation}
where the corresponding $\nu^2$ is a solution of the equation
\begin{equation}
\Xi \sum_j \frac {k_j^2\omega_j ^3}{\omega_j^2-\nu^2}=1 \ .
\label{angfreq:eqn}
\end{equation}
As a simple graphical analysis suggests, this equation has as many
solutions $\nu_j^2$ as original modes.  For any frequencies $\omega_j$ and
couplings $k_j$, the lowest solution is $\nu_1=0$, corresponding to an
eigenvector $x_i=(1,1,1,...)$, i.e.\ to the totally symmetrical coordinate
($\sum \theta_i/n$): it represents the ``soft mode'' of pseudo-rotation
along the circular JT valley.  If the other solutions $\nu_j$ are sorted in
ascending order of frequency, the inequality $\omega_{j-1}\leq \nu_j\leq
\omega_j$ can be seen to hold.  In the special case of modes of the same
frequency $\omega_{j-1}=\omega_j$, the implied equality $\nu_j=\omega_j$ is
also true.  Indeed, a set of $d$ modes with equal frequencies can be
rewritten as a single coupled mode, plus $d-1$ uncoupled ones: the
unchanged frequencies $\nu_j=\omega_j$ correspond therefore to the angular
part of the uncoupled modes.  In the simple, but instructive, case of $n=2$
modes, the (single) nonzero angular frequency is
\begin{equation}
\label{nu2:eqn}
\nu_2=\left[ \omega_1 \omega_2 (k_2^2
\omega_1+k_1^2 \omega_2)/(k_1^2 \omega_1+k_2^2 \omega_2)\right]^{1/2}\ .
\end{equation}
This expression for the new angular frequency has the form of a weighted
geometrical mean between the unperturbed frequencies, with such weights
that $\nu_1$ is attracted towards the frequency of the mode with weaker
coupling. This is reasonable because the frequency of any mode whose
coupling is exactly zero should of course remain exactly unchanged. The
frequencies $\nu_j$ display the same behavior in the general case: the new
modes $\nu_j$ are located in the intervals between neighboring frequencies
$\omega_{j-1}$, $\omega_j$, attracted towards the mode with {\em weaker}
coupling strength $k_j$.  In the limiting case $k_j=0$, therefore, $\nu_j$
restores smoothly the twofold-degeneracy of the uncoupled mode $\omega_j$.
For a point impurity coupled to the phonon continuum, equations analogous
to Eq.\ (\ref{angfreq:eqn}) were derived earlier\cite{Bersuker}, but apparently
not put to practical use.  Here, we will use our eigenvalue equation
(\ref{angfreq:eqn}) for the calculation of vibron frequencies, and of the
associated zero-point energy.

We underline, incidentally, that this equation has
a very familiar form, namely that of a Slater-Koster scattering problem for
a separable potential, or of a collective-mode equation\cite{Fano92},
or of a BCS gap equation\cite{Schrieffer}.  By
coupling to the same electronic state, the different vibrational modes get
effectively coupled to each other by a kind of separable ``attractive
$\delta$-function'' potential. The lowest bound state is forced to zero
frequency, $\nu_{1}= 0$, by what appears to be an exact sum rule, related
to the O(2) symmetry of the Hamiltonian (\ref{hamiltonian:eqn}),
corresponding to the independence of the energy of the angle $\theta$.  In the
BCS case, broken gauge invariance and Goldstone's theorem give rise to a
formally similar sum rule. A universal feature of linear JT systems,
the zero mode is also present in the phonon continuum case\cite{Bersuker}.

In the evaluation of the zero-point energy, the normal angular
frequencies appear only as a sum:
\begin{eqnarray}
\label{Emmodes:eqn}
E_{\rm JT} - E(k_i=0) &=&
 - \frac 12 k_{\rm eff}^2 \omega_{\rm eff} + \frac 12 \left(
\sum_i \omega_i + \sum_{{\rm angular~modes~}\nu_j} \nu_j \right)
- \sum_i \omega_i +O(k_{\rm eff}^{-2}) \nonumber\\
&=&
 - \frac 12 k_{\rm eff}^2 \omega_{\rm eff} + \frac 12 \sum_j (\nu_j -\omega_j)
	+O(k_{\rm eff}^{-2}) \ .
\end{eqnarray}
This main result is not in ``closed form'', since the $\nu_j$ are defined
implicitly as the solutions of Eq.~(\ref{angfreq:eqn}).  However this is a
very marginal shortcoming, involving in each specific case the simple
solution of a numerical equation, explicitly given in terms of the
parameters $\omega_j$ and $k_j$.  We also remark that a rescaling of all
the couplings $k_j\longrightarrow \alpha k_j$ brings no change to the
secular equation (\ref{secular:eqn}): the angular frequencies $\nu_j$,
therefore, do not depend on $k_{\rm eff}$, but only on the ratios among the
couplings. Thus, in particular, the $\sum \nu_j - \omega_j$ zero-point term
is really a term of order zero in $k_{\rm eff}$, as it should.

\begin{figure}
\epsfxsize 10.0cm
\inseps{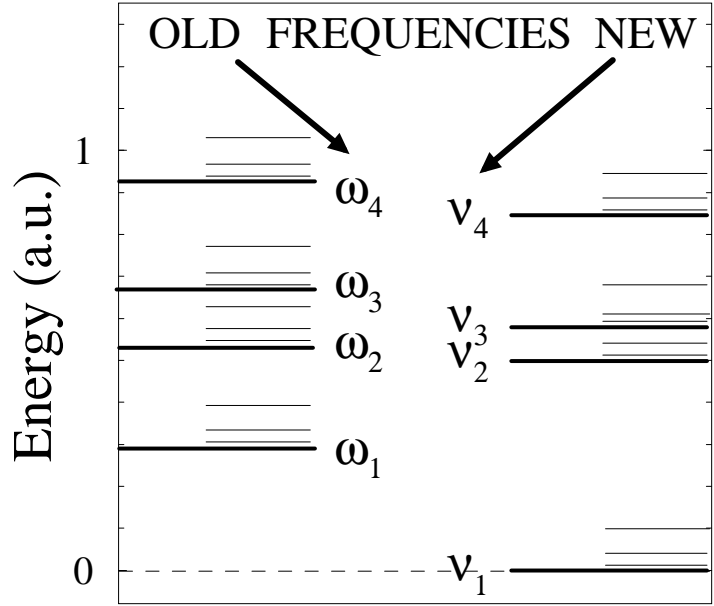}{1.0}
\caption{ A schematic picture of the spectrum of vibronic levels of a
coupled many-modes JT system, in the strong-coupling limit.  For
simplicity, only the harmonic one-phonon (fundamental) states are drawn.
The thin lines represent a few low-lying pseudorotational levels.  In the
$e\otimes (n\ E)$ case the harmonic vibrations are nondegenerate, but all
the states acquire a twofold degeneracy when the pseudorotation is
considered.  The $\omega_i$ and $\nu_i$ fundamentals in $t\otimes (n\ H)$
are threefold and twofold degenerate, respectively, and these degeneracies
combine further with those of the ``soft'' states of pseudorotation around
the trough.
\label{spec:fig}}
\end{figure}

This approach provides quite naturally the detailed structure of the
spectrum of the excited states in the strong-coupling many-modes case. All
the uncoupled frequencies $\omega_i$ persist as radial frequencies in the
DJT spectrum, while new harmonic angular modes of intermediate frequencies
$\nu_j$ (solutions of Eq.~(\ref{angfreq:eqn})) appear in between, as
illustrated schematically in Fig.~\ref{spec:fig}.

In addition to these main spectral structure, the zero-frequency mode
$\nu_1$ introduces an entire ladder of low-energy excitations,
corresponding to the quantization of the pseudo-rotation around the
manifold of minimum BO potential.  We write these excitations in terms of
the ``moment of inertia'' of the pseudo-rotor associated with the $\sum
\theta_i$ coordinate\cite{Bersuker}:
\begin{equation}
\label{erot:eqn}
E^{\rm rot}(m)=\frac {1}{2 \sum k_i^2 \omega_i^{-1}} \cdot m^2
\end{equation}
where the allowed values of $m$ are $m=\pm\frac 12, \pm\frac 32,...$, as
required by the Berry-phase prescription.  $E^{\rm rot}(m)$ adds to $E_{\rm
JT}$ to give the low-energy states.  In particular, the energy of the
(twofold degenerate) ground state is given in this scheme by
\begin{equation}
\label{egsrot:eqn}
E_{\rm GS}\approx E_{\rm JT} + E^{\rm rot}\left(\pm\frac 12\right)\ .
\end{equation}

\subsection{Numerical calculations for $e\otimes (n E)$ }
\label{numerical:sect}

The relative accuracy of the results just derived and of O'Brien's
effective-mode approximation\cite{ob72}
\begin{eqnarray}
\label{Emmobrien:eqn}
E_{\rm JT,eff}^{(0)} - E_{\rm JT}(k_i=0) &=&
 - \frac 12 k_{\rm eff}^2 {\omega_{\rm eff}}
- \frac 12 \omega_{\rm eff} - \frac 14 \frac {{\overline {\omega^2}} -
	{\omega_{\rm eff}}^2}{\omega_{\rm eff}} + O(k_{\rm eff}^{-2})
\nonumber\\
&=& - \frac 12 k_{\rm eff}^2 {\omega_{\rm eff}}
	- \frac 14 \frac {{\overline {\omega^2}} +
		{\omega_{\rm eff}}^2}{\omega_{\rm eff}} + O(k_{\rm eff}^{-2})
\end{eqnarray}
where ${\overline {\omega^2}}=\sum_i k_i^2 \omega_i^2/k_{\rm eff}^2$, can
be assessed by comparison with the exact numerical diagonalization results of
(\ref{hamiltonian:eqn}) on a truncated oscillator basis.  The ground-state
energy obtained by numerical diagonalization is a variational estimate of
the exact ground-state energy.  It converges very rapidly as the number $N$
of included oscillator states exceeds $k_{\rm eff}^2$.  In practice, for
$k_{\rm eff}^2\approx 100$, inclusion of $N=100$ quanta in the oscillators
ladder yields an accuracy of order $10^{-8}\omega_{\rm eff}$, largely
sufficient for our purposes.

For the case of only two modes, where an explicit form
(\ref{nu2:eqn}) is available
for $\nu_2$, the difference between the two is
\begin{equation}
\label{diff2modes:eqn}
E_{\rm JT,eff}^{(0)}-E_{\rm JT} =
	- \frac 14 \frac {{\overline {\omega^2}} +
		{\omega_{\rm eff}}^2}{\omega_{\rm eff}} +
	\frac 12 (\omega_1+\omega_2-\nu_2)
	+O(k_{\rm eff}^{-2}) \ .
\end{equation}

As a first example, we consider the case of two modes with similar
frequencies, such that $\omega_2- \omega_1=\delta \omega$ is a small
parameter. The difference in ground-state energy give by the two
approximations is
\begin{equation}
\frac{E_{\rm JT,eff}^{(0)}-E_{\rm JT} }{\omega_{\rm eff}} \approx
\left(\frac {k_1 k_2 ~ \delta \omega}{2 \omega_{\rm eff} k_{\rm eff}^2}
\right)^4
+ 2 (k_2^2 -k_1^2) (k_1 k_2)^4
\left(\frac {\delta \omega}{2 \omega_{\rm eff} k_{\rm eff}^2}
\right)^5 +...
\end{equation}
This difference is therefore very small for small $\delta \omega$: in this
limit the two expressions are essentially coincident, and we verified that
they both agree with the exact numerical ground-state energy at strong
coupling.  This is not surprising, since we know that the effective-mode
result is exact in the trivial case of equal
frequencies\cite{Bersuker}, and it confirms that also our expansion
is correct in this limit.

\begin{figure}
\epsfxsize 10.0cm
\inseps{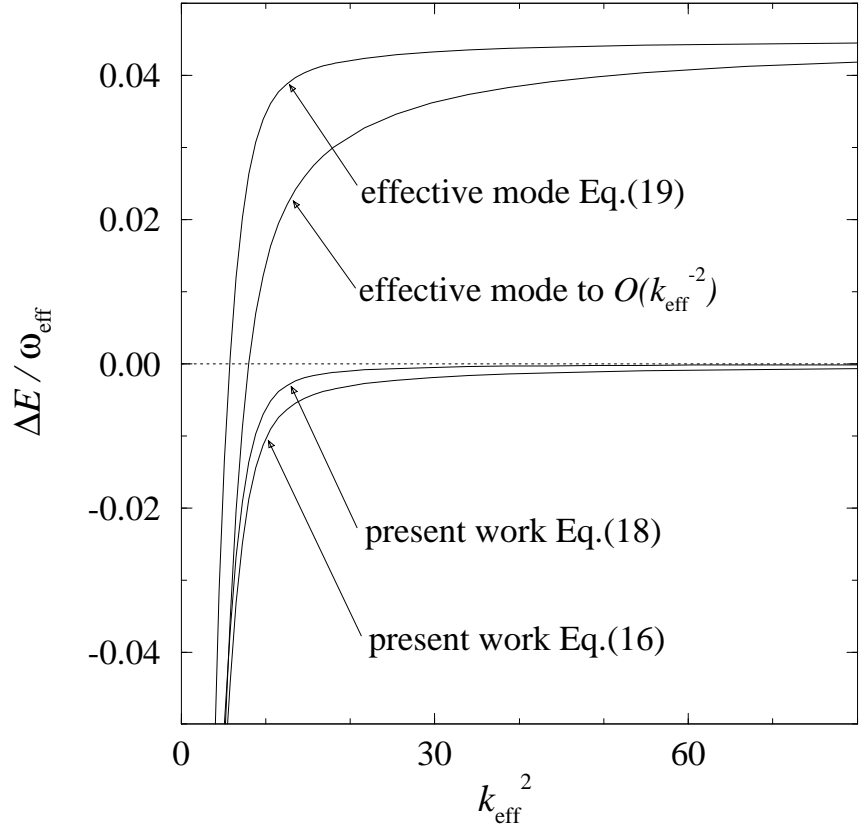}{0.8}
\caption{ The difference between the approximate expressions and the exact
ground-state energy, obtained by Lanczos diagonalization including up to
$N=100$ oscillator states, for two modes of frequencies $\omega_2=10\cdot
\omega_1$, as a function of the total JT coupling strength $k_{\rm
eff}^2=2k_1^2=2k_2^2$.  The figure shows the semiclassical expression
(\protect\ref{Emmodes:eqn}), its correction including the pseudo-rotation
contribution (\protect\ref{egsrot:eqn}), and the effective-mode formula
(\protect\ref{Emmobrien:eqn}), with also the complete version -- Eq.~(44)
of Ref.~\protect\cite{ob72} -- including the $k_{\rm eff}^{-2}$ correction.
\label{2modes:fig}}
\end{figure}

Moving to the more interesting situation of very different
frequencies, for example $\omega_2=10\cdot \omega_1$, we find that
the correction becomes more important ($\sim 5\%~\omega_{\rm eff}$).
Fig.~\ref{2modes:fig} shows  the error of the approximate expressions
$\Delta E = E^{\rm approx}-E^{\rm exact}$ for the ground-state energy at
different values of the coupling.  Our approximate formula
(\ref{Emmodes:eqn}) converges systematically (from below) to the exact
energy modulo corrections $\sim k_{\rm eff}^{-2}$.  The effective mode
expression (\ref{Emmobrien:eqn}) differs from Eq.~(\ref{Emmodes:eqn}) for a
quantity depending on the individual frequencies and on the ratio between
the couplings, but {\em not} on the total coupling strength $k_{\rm
eff}^{-2}$.  This difference introduces a systematic shift in the
strong-coupling limit of the the expression (\ref{Emmobrien:eqn}).

In the extreme case of very small frequency ratio $\omega_1/ \omega_2 \ (<<
\left[ \frac{k_1}{k_2} + \frac{k_2}{k_1}\right]^{-2})$, the systematic
shift introduced by Eq.~(\ref{Emmobrien:eqn}) becomes as large as
\begin{equation}
\frac{E_{\rm JT,eff}^{(0)}-E_{\rm JT} }{\omega_{\rm eff}} \approx
\frac 14 \left(\frac{k_1}{k_2}\right)^2
- \frac 12 \left[\frac{k_1}{k_2} + \left(\frac{k_1}{k_2} \right)^3\right]
\left(\frac{\omega_1}{\omega_2} \right)^{1/2}
+ \frac 12 \left[1+\left(\frac{k_1}{k_2} \right)^2 \right]
\frac{\omega_1}{\omega_2}
+... \ ,
\end{equation}
which is a relevant fraction of $\omega_{\rm eff}$.

\begin{figure}
\epsfxsize 10.0cm
\inseps{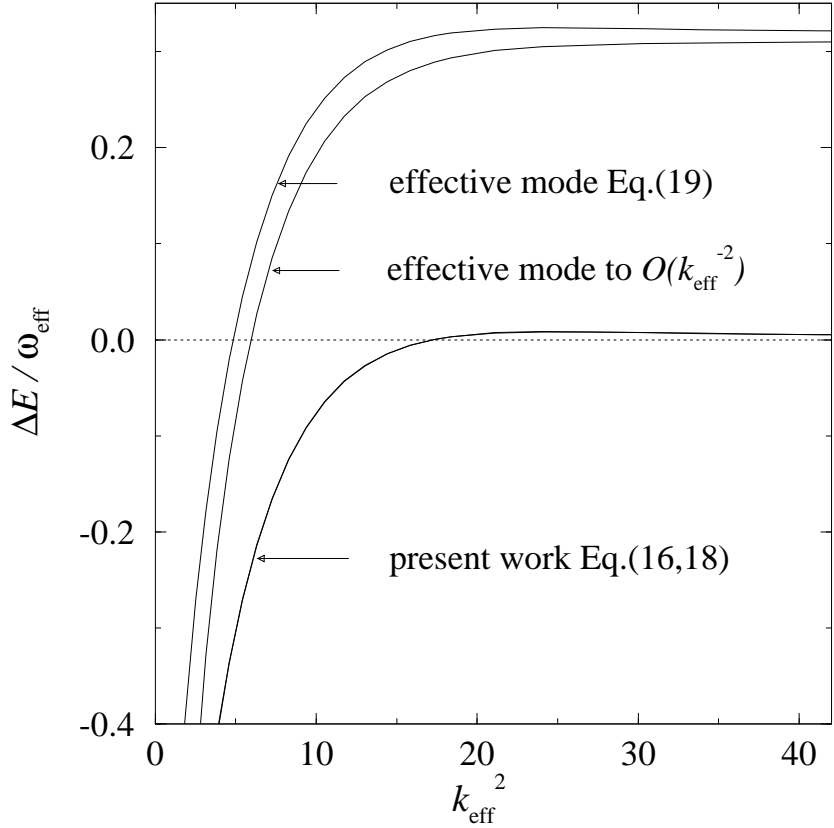}{0.8}
\caption{ The difference between the approximate expressions and the exact
ground-state energy, obtained by Lanczos diagonalization including up to
$N=40$ oscillator states, for $n=3$ modes of frequencies
$\omega_1=\omega_2/100=\omega_3/120$, and couplings $k_1=2 k_2=2 k_3$, as a
function of $k_{\rm eff}^2$.  The semiclassical expression
(\protect\ref{Emmodes:eqn}) and its correction including the
pseudo-rotation contribution (\protect\ref{egsrot:eqn}) are
indistinguishable on the scale of the figure.  The figure shows also
$\Delta E$ for the perturbative formula (\protect\ref{Emmobrien:eqn}), and
for the version including the $k_{\rm eff}^{-2}$ correction.
\label{3modes:fig}}
\end{figure}

A third example (Fig.~\ref{3modes:fig}) illustrates the validity of the
method for the case of three modes.  For a rather large frequency spread
$\frac {{\overline {\omega^2}} - {\omega_{\rm eff}}^2}{\omega_{\rm
eff}^2}(\sim 2)$, the approximate formula (\ref{Emmodes:eqn}) still
converges to the exact energy, but not monotonically in $k_{\rm eff}$.

An interesting observation suggested by Figs.~\ref{2modes:fig} and
\ref{3modes:fig} is that, for weak enough coupling -- say $k_{\rm
eff}\lesssim 3$, the effective-mode theory, owing to the systematic shift
discussed above, can yield {\em better agreement} with the exact
ground-state energy than our method, which is instead
superior at strong coupling.  Thus the effective-mode method and the
present one are to some extent complementary.

\begin{figure}
\epsfxsize 10.0cm
\inseps{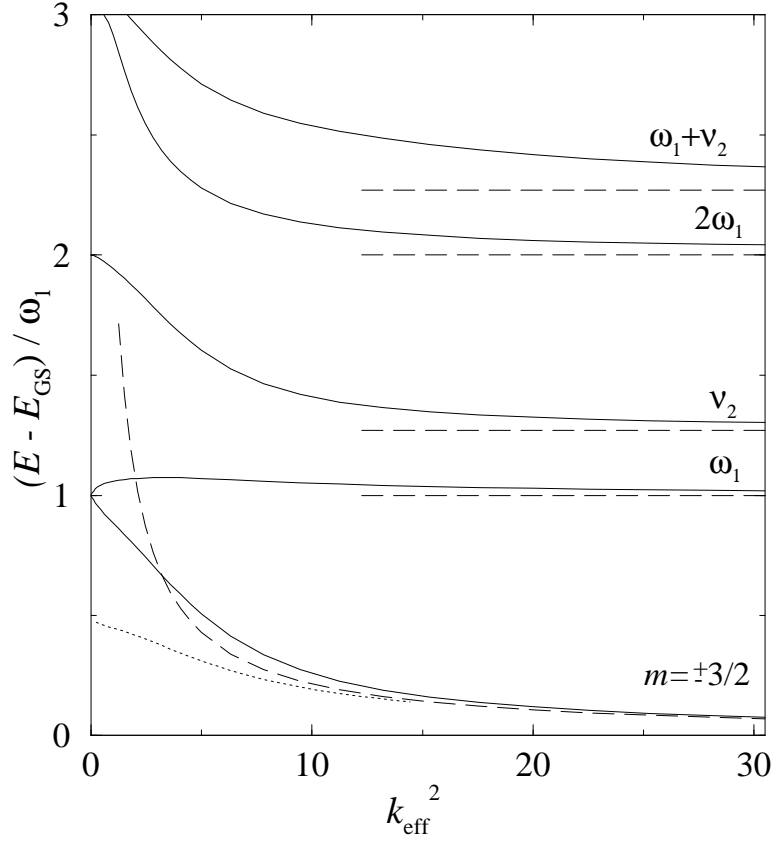}{0.8}
\caption{ Excitation energy (solid lines) of selected low-lying excited
states for $n=2$ modes of frequencies $\omega_1=1$, $\omega_2=3$, and
coupling ratio $k_1=\frac 12 k_2$ (solid lines).  The four lowest $m=\pm
\frac 12$ states and the lowest $m=\pm \frac 32$ state are drawn, and
compared with the theoretical previsions (dashed lines) and assignments of
our model, valid at strong coupling.  The dotted line represents the
excitation energy of the lowest $m=\pm \frac 32$ state including the
self-consistent centrifugal correction of Ref.~\protect\cite{Bersuker}.
\label{exc:fig}}
\end{figure}

Besides the ground state, the Lanczos technique employed for
diagonalization easily generates a few low-lying excited states.  These
accurate excitation energies can be compared with our approximate
frequencies $\nu_i$'s.  In Fig.~\ref{exc:fig} we report the excitation
energy of the
low-lying $m$=$\pm\frac 12$ states.  These energies show a clear
convergence to the fundamentals, overtones and combination states
of $\omega_1$ and
$\nu_1$, as expected from the theory.  Above each of these origins,
higher-$m$ states form pseudorotational ladders, as determined by $E^{\rm
rot}(m)$.  In Fig.~\ref{exc:fig} we represent for clarity only the
$m=\pm\frac 32$ excitation above the ground state, and compare it with the
theoretical value given by Eq.~(\ref{erot:eqn}).  Interestingly, a
parallel comparison
with the effective-mode model shows that, even though the ground-state
energy estimates in the two models differ, the values of the
pseudorotational quantum of energy (expressed in completely different
forms) are numerically coincident\cite{pseudosum:note}.

We stress that our adiabatic potential (\ref{BOpotential:eqn}) does not
explicitly include centrifugal terms, unlike what is customarily done away
from strong coupling.  Reference~\cite{Bersuker}, for example, describes a
self-consistent prescription for embodying these terms at the outset.  For
the sake of completeness we include in Fig.~\ref{exc:fig} the excitation
energy to the first $m=\pm\frac 32$ pseudorotational state obtained through
that method. The self-consistent procedure renormalizes the inertial
moment, eliminating the $k_{\rm eff}^{-2}$ divergence at weak coupling, but
it ends up providing a worse approximation to the exact energy in the
region of intermediate to strong coupling. Hence, inclusion of centrifugal
terms does not appear to be useful in this regime.
Concerning the $O(k_{\rm eff}^{-2})$ terms, we note that, while in
Fig.~\ref{2modes:fig} the pseudorotational contribution $E^{\rm
rot}\left(\pm\frac 12\right)$ shifts the approximate ground-state energy
closer to the exact one, in Fig.~\ref{3modes:fig} for $k_{\rm eff}\gtrsim
4$ the error of the approximate expression (\ref{Emmodes:eqn}) is {\em
positive}: thus addition of the pseudorotational contribution moves the
approximation further away from the correct value.  We conclude that the
pseudorotational contribution as given by Eq.~(\ref{erot:eqn}), contrary to
the single-mode case, does {\em not} exhaust all the $O(k_{\rm eff}^{-2})$
corrections to the truncated expansion (\ref{Emmodes:eqn}).
We will not further investigate such terms in the present work.

\section{Many modes $t\otimes (n\ H)$}
\label{txH:sect}

After the basic $e\otimes (n E)$ system, treated in the previous sections,
we wish now to consider a second, different case, to illustrate how much of
the procedure used can be carried over.  The interaction of an orbital
triplet ($t$) with a set of fivefold vibrational modes ($H$ icosahedral
representations) is our next choice.  As we shall see, apart from some
technical differences, we will be able to follow very closely the approach
that proved successful for the $e\otimes (n\ E)$ case.

We start with the Hamiltonian operator\cite{AMT,ob96}
\begin{eqnarray}
\label{hamiltoniant:eqn}
H &=& \frac 12 \sum_i \omega_i \left(\vec p_i^2 + \vec q_i^2 \right) +
\frac 12 \sum_\sigma \\
&&	\left(c_{x\sigma}^\dagger, c_{y\sigma}^\dagger,
		c_{z\sigma}^\dagger \right)
	\sum_i k_i \omega_i
\pmatrix{
q_{i1}-\sqrt{3}q_{i4} &-\sqrt{3}q_{i3}		&-\sqrt{3}q_{i2} \cr
-\sqrt{3}q_{i3}	&q_{i1}+\sqrt{3}q_{i4}	&-\sqrt{3}q_{i5}\cr
-\sqrt{3}q_{i2}	&-\sqrt{3}q_{i5}	& -2q_{i1} }
                                        \pmatrix{       c_{x\sigma}\cr
                                                        c_{y\sigma}\cr
                                                        c_{z\sigma} }
\nonumber
\end{eqnarray}
where now the three degenerate electronic states are labeled $x,y,z$, and
$\vec q_i$ indicate a five-dimensional vector of components $\left(
q_{i1},... q_{i5}\right)$.  Reference~\cite{ob96} introduces a polar
parametrization of the five-dimensional space of one single mode $i$ in
term of a radial coordinate $q_i$, plus four angles $\alpha_i$, $\theta_i$,
$\phi_i$, and $\gamma_i$.  The same work\cite{ob96} reports for a single
mode $i$, the expression for the lowest electronic eigenvalue of the
one-electron matrix in terms of these coordinates:
\begin{equation}
- \cos(\alpha_i) k_i \omega_i q_i \ .
\end{equation}
The BO potential (obtained adding $\sum q_i^2/2$ to the electronic part) is
therefore independent of $\theta_i$, $\phi_i$, and $\gamma_i$, but it
depends explicitly on the angle $\alpha_i$.  In particular, it is minimum
at $q_i=k_i$, $\alpha_i=0$.  For this special value, the angular
parametrization of $\vec q_i$ becomes singular, the coordinate $\vec q_i$
being independent of the angle $\gamma_i$.  The DJT valley of $t\otimes (n\
H)$ is thus, as well known, two-dimensional, parametrized by $\theta_i$,
$\phi_i$.

Now, let us move on to $n>1$ fivefold modes.  Consider the vector $\sum_i
k_i \omega_i \vec q_i$ intervening in the JT coupling matrix
(\ref{hamiltoniant:eqn}): if we indicate with $\alpha$, $\theta$, $\phi$,
and $\gamma$ the corresponding set of polar angles, the lowest electronic
eigenvalue for the general case writes
\begin{equation}
- \cos(\alpha) \left|\sum_ik_i \omega_i \vec q_i\right|=
- \cos(\alpha) \left(\sum_i k_i^2 \omega_i^2 q_i^2
+ 2 \sum_{i<j} k_i k_j \omega_i \omega_j {\vec q_i} \cdot {\vec q_j}
\right)^{1/2} \ .
\end{equation}
The minimization of the corresponding the BO potential, function of $5n$
variables, is therefore very similar to the analogous for
Eq.~(\ref{BOpotential:eqn}): here the electronic term is expressed as the
negative of a product, where both terms can be maximized at the same time.
As a result, at the BO minimum, $\alpha=0$, and all the $\vec q_i$ should
align with one another in a common direction:
\begin{eqnarray}
\label{minimat:eqn}
q_i^{\rm min}&=&k_i\nonumber \\
\alpha_i^{\rm min}&=&\alpha=0\nonumber \\
\theta_i^{\rm min}&=&\theta\\
\phi_i^{\rm min}  &=&\phi\nonumber \\
\gamma_i^{\rm min}&=&\gamma\nonumber
\end{eqnarray}
Again the trough is two dimensional, parametrized by $\theta$, $\phi$,
since $\gamma$ is singular as discussed in the one mode case above.  The
expression (\ref{Eclass:eqn}) for the classical JT energy gain holds in the
$t\otimes (n\ H)$ case.

The next step is to generate the quantum corrections to the classical
result.  For this purpose we need the harmonic
frequencies of oscillation  around the minimum (\ref{minimat:eqn}).  Here a
completely
analytical approach, as in Sec.~\ref{exE:sect}, fails, for two reasons.  As
a first point, the factor $\cos(\alpha)$, contributing to the Hessian
matrix $\frac{\partial^2 V}{\partial x_\mu \partial x_\nu}$, is unavailable
as an explicit expression in terms of the coordinates $\vec x=\left(
q_1,q_2,..., \alpha_1, \alpha_2,..., \theta_1,....\right)$.  Secondly, the
parametrization is singular right at the minimum, thus it cannot generate
all the angular modes.

We resort therefore to an alternative, more numerical approach.  We express
explicitly the lowest electronic eigenvalue in {\em Cartesian} coordinates
$\vec x=(q_{11},..., q_{15},q_{21},..., q_{n5})$ compute
the $5n\times 5n$ Hessian matrix (at an arbitrary minimum point in the
trough) and the kinetic matrix [see Eq.~(\ref{harpoteq:eqn}),
(\ref{kineq:eqn})], and diagonalize the product ${\bf T}\left(\vec{x}^{\rm
min}\right)^{-1} \cdot {\bf V}$ [see Eq.~(\ref{dinmat:eqn})].  For $n=2$,
the calculation can be done analytically, choosing some special points in
the minimum manifold (\ref{minimat:eqn}), such as $\theta=0$,
$\phi=$anything [i.e.\ $\vec q_i=(k_i,0,0,0,0)$], or $\theta=\pi/4$,
$\phi=\pi/2 $ [i.e.\ $\vec
q_i=(k_i/4,0,0,-k_i\sqrt{3}/4,k_i\sqrt{3}/2)$]. The resulting matrix has
two threefold-degenerate eigenvalues of frequency $\omega_1$ and $\omega_2$
respectively, a twofold eigenvalue $\nu_1=0$, and a second twofold
eigenvalue of frequency $\nu_2$ given by Eq.~(\ref{nu2:eqn}).  The
eigenvalues are of course independent of the choice of the minimum point
around which the expansion is done.  In the general case of larger $n$, we
diagonalized numerically the dynamical matrix, obtaining for any choice of
the frequencies and couplings $n$ threefold-degenerate eigenvalues
$\omega_1$,... $\omega_n$, and $n$ new twofold-degenerate eigenvalues,
which we called $\nu_i$.  Again we have always $\nu_1=0$, and all $\nu_i$'s
are located between subsequent modes $\omega_{i-1}$ and $\omega_i$.  Thus,
directed by the similarity with the $e\otimes (n\ E)$ case, we verified by
substitution that the new frequencies $\nu_i$ are solutions of
Eq.~(\ref{angfreq:eqn}).

 From the above analysis we conclude therefore that (i) in analogy to the
$e\otimes (n\ E)$ case, each original frequency $\omega_i$ is still present
in the e-v coupled spectrum; (ii) in $t\otimes (n\ H)$ each $\omega_i$
corresponds not just to the radial degrees of freedom, but includes pairs
of coupled
angular modes as well, for a total degeneracy of three; (iii) new harmonic
vibron modes $\nu_i$ appear at the same frequencies -- dictated by the same
equation (\ref{angfreq:eqn}) -- as in the $e\otimes (n\ E)$ case (see
Fig.~\ref{spec:fig}); (iv) in the $t\otimes (n\ H)$ case, the new
modes $\nu_i$ are all twofold degenerate; (v) in particular, the lowest new
frequency $\nu_1=0$ again corresponds to the free pseudorotation around the
minimum trough; (vi) finally, the semiclassical DJT energy gain, in analogy
to Eq.\ (\ref{Emmodes:eqn}), is for this case:
\begin{eqnarray}
\label{Emmodest:eqn}
E_{\rm JT} - E(k_i=0) &=&
 - \frac 12 k_{\rm eff}^2 \omega_{\rm eff} + \frac 12 \left(
3 \sum_i \omega_i + 2 \sum_{j} \nu_j \right)
- \frac 52 \sum_i \omega_i +O(k_{\rm eff}^{-2}) \nonumber\\
&=&
 - \frac 12 k_{\rm eff}^2 \omega_{\rm eff} + \sum_j (\nu_j -\omega_j)
	+O(k_{\rm eff}^{-2}) \ .
\end{eqnarray}

\subsection{C$_{60}^-$: a case of intermediate coupling $t\otimes (n\ H)$}

As an application, chosen to test the limits of applicability of our
approximation, we consider the DJT problem of the fullerene
anion\cite{AMT,MTA,Ihm,Reno}, where a single electron in a $t_{1u}$
electronic state couples to $n=8$ $H_g$ vibrational modes. The frequencies
and couplings of the eight modes differ strongly, and none of the couplings
is particularly large.
The numerical values of the parameters for the C$_{60}^-$ ion, the same
adopted in our previous work\cite{Reno}, are reported in
Table~\ref{param:table}.  The table gives also the new vibron frequencies
$\nu_j$ of the the coupled system, calculated within our approximation.
Note that mode~6 does not give rise to a new frequency, since its coupling
is vanishingly small.  On the basis on these numbers, we have computed the
ground-state energy gain in the approximation (\ref{Emmodest:eqn}), and we
report it in Table~\ref{GS:table}, along with the corresponding values in
the effective-mode scheme, as presented in Ref.~\cite{ob96}, and with the
exact result, obtained by Lanczos diagonalization\cite{Reno,Gunnarsson}.

We note that the two approximations, the effective mode and ours, are
essentially equivalent, both of them in error by some 25\% relative to the
exact result. The approximate energy gains, in particular, correct in
excess the initial classical JT gain $\left| E_{\rm class} \right|$ (too
small, by about a factor $1/2$) and are now 25\% too large.  The reason for
this is that the vibron zero-point energy, here larger than the classical
energy gain itself, overestimates the true quantum correction, indicating
important higher-order corrections.  This reflects the fact that, in
C$_{60}^-$, $k_{\rm eff}$ is only slightly larger than unity, since the
individual couplings are rather weak, so that a semiclassical expansion
truncated omitting terms of $O(k_{\rm eff}^{-2})$($\approx250$cm$^{-1}$)
and higher is quite far from convergence.  Qualitatively, this regime
corresponds to the region $k_{\rm eff}^2\approx 2.5$ in
Fig.~\ref{3modes:fig}.

In the detail, the effective-mode energy is a few wave numbers better than
our approximation. The reason for that is the asymptotic shift previously
discussed for the $e\otimes (n\ E)$ case, and illustrated in
Figs.~\ref{2modes:fig} and \ref{3modes:fig}, which makes the effective mode
competitive when the coupling is weaker.

If in addition we consider the contribution, of order $k_{\rm eff}^{-2}$,
of pseudo-rotations, we can translate Eq.~(\ref{erot:eqn}) to:
\begin{equation}
\label{erott:eqn}
E^{\rm rot}(L)=\frac {1}{6 \sum k_i^2 \omega_i^{-1}} \cdot L~(L+1)\ .
\end{equation}
The inclusion of this contribution (60.1cm$^{-1}$) for the $L=1$ ground
state\cite{AMT,ob96} brings the semiclassical estimate of the ground-state
energy gain to within 20\% of the exact numerical value
(Table~\ref{GS:table}).  Note however that, according to (\ref{erott:eqn}),
the first-excited ($L=3$) pseudorotational state should be found at
$\sim$300cm$^{-1}$ above the ground state, that is slightly {\em above}
$\omega_1$.  This confirms that the fullerene ion is really in an
intermediate-coupling situation, rather far from all approximate limiting
regimes.

\section{Discussion}

O'Brien's approach -- based on an ``effective-mode'' picture -- is
available in the literature for the description of the low-energy states of
a many-modes JT system.  It introduces a fictitious, effective single mode,
treating perturbatively the residual corrections.  This approach,
conceptually attractive as it is (and exact in the limit of equal
frequencies $\omega_i=\omega$)\cite{Bersuker}, is not completely
satisfactory, particularly when the spread of the frequencies is sizable,
and couplings are large. In that case, the effective-mode expression
(\ref{Emmobrien:eqn}) for the ground-state energy gain has a deviation
which may be a large fraction of the effective frequency $\omega_{\rm
eff}$.  This shift, for given frequencies and ratios among the couplings,
is independent of $k_{\rm eff}$, therefore the inclusion of higher power
corrections in $k_{\rm eff}^{-2}$ to Eq.\ (\ref{Emmobrien:eqn}), such as those
of Eq.~(44) of Ref.~\cite{ob72} cannot correct the asymptotic behavior at
large $k_{\rm eff}$.

We have introduced a semiclassical treatment of the many-modes problem, which
corrects the difficulties of the effective-mode model.  We have applied this
method to two specific JT cases making it clear that,
with straightforward modifications, it can be
extended to other DJT systems.  The method is based on the approximate
calculation of the new vibron normal mode frequencies $\nu_i$
arising after DJT coupling. Of these modes,  $n-1$ fall in the
intervals between successive original bare mode
frequencies, the remaining one is a zero mode $\nu_1=0$.
The approximate $\nu_i$'s are again
solutions of a collective-mode--type equation\cite{Fano92}, already
obtained in an equivalent form in previous studies
of the continuum case\cite{Bersuker}.
Numerical tests confirm this picture, including the recovery of the
original, {\em unperturbed} frequencies $\omega_i$'s in the strong-coupling
spectrum.

In particular, we obtain expressions
(\ref{Emmodes:eqn},\ref{Emmodest:eqn}), in terms of these frequencies, for
the ground-state energy which are exact in the strong-coupling limit.  At
finite coupling, our formulas introduce an error of the order $k_{\rm
eff}^{-2}$, which is only partly corrected by the term Eqs.\
(\ref{erot:eqn}),(\ref{erott:eqn}) of pseudorotation along the BO minimum
trough with the associated Berry phase constraints\cite{AMT}.
However, in cases with strong enough coupling for our
expansion to hold for the potential, the
expressions to order $O(k_{\rm eff}^0)$ gives satisfactory results, and
will not need further refinement.

We provide ample numerical verification for our expectations in the
$e\otimes (n\ E)$ case. A detailed numerical test of the result for the
case of $t\otimes (n\ H)$ would be rather difficult, because of the
explosion of the basis set size, which occurs when many vibron states
are included, as necessary at strong coupling.  The
case of C$_{60}^-$, not a very strongly-coupled system, therefore
amenable to exact diagonalization, shows that
our approach provides results for the energy gain which are basically
equivalent to those of the effective-mode approximation.

More interesting would be a direct experimental confirmation of the new
vibron frequencies which we have found. The ideal test system could be a
molecular system where the couplings to a few (2 or 3) modes are known,
reasonably strong, controlled, and possibly switchable on and off (for
example upon doping): in such a situation, infrared/Raman spectroscopy
should easily evidence the birth of new modes.  Benzenoid cations might
represent a suitable systems for such future investigations.

The present formulation of course neglects a large set of effects
potentially capable of modifying the picture to some degree, especially on
the low- and high-energy ends of the vibronic spectra.  The most obvious
limitation regards the assumed harmonicity of the ``small oscillations''
(\ref{BOpot:eqn}), and the neglect
from the beginning of nonlinear terms in (\ref{hamiltonian:eqn}).  In
fact, there is no truly flat JT valley in a real system.  The effect of valley
warping is to localize the pseudorotational motion into a discrete set of
minima.  The importance of this phenomenon is determined by
the relative value of the pseudorotational energy $E^{\rm rot}$ and 
the height of the barriers between minima.
Initially the warping will affect mostly the zero-energy mode, but
for very large JT distortions, the whole shape of the BO surface will
change.  As a consequence, our results apply best to a
strong-coupling case where the large value of $k_{\rm eff}$ is given by the
addition of many individually small contributions $k_i$, such that the
distortion of each mode's coordinate remains small, and the effect of
higher-order terms is therefore weak.

\section{Acknowledgements}

We are thankful to Paolo De Los Rios and Fabio Pistolesi for useful
discussions. We also acknowledge financial support from INFM,
projects LOTUS and HTSC, and from the European Commission, through HCM
contract ERBCHRXCT940438, and TMR FULPROP project.

\begin{table}
\begin{center}
\begin{tabular}{crrrr}
$i$	&$\omega_i$(cm$^{-1}$)&	$k_i$	&$\frac 12 k_i^2 \omega_i$(cm$^{-1}$)&
$\nu_i$(cm$^{-1}$)\\
\hline
1	&	270.0	&	0.868	&	101.71 	&	0.0	\\
2	&	430.5	&	0.924	&	183.78 	&	329.5	\\
3	&	708.5	&	0.405	&	58.11 	&	633.8	\\
4	&	772.5	&	0.448	&	77.52 	&	742.3	\\
5	&	1099.0	&	0.325	&	58.04 	&	1031.8	\\
6	&	1248.0	&	0.000	&	0.00 	&	1248.0	\\
7	&	1426.0	&	0.368	&	96.56 	&	1302.4	\\
8	&	1575.0	&	0.368	&	106.65 	&	1519.0	\\
	&$\omega_{\rm eff}=581.1$ & $k_{\rm eff}=1.532$ &
$E_{\rm class}=\sum \frac 12 k_i^2 \omega_i=682.36$	&	\\
\end{tabular}
\end{center}
\caption{The frequency/coupling parameters for many-modes DJT of
C$_{60}^-$.  The last line combines them to give the effective parameters
introduced in Ref.\protect\cite{ob72}.  With $k_{\rm eff}^2=2.35$,
C$_{60}^-$ is indeed an intermediate-coupling system.  The last
column reports the new frequencies of the DJT-coupled system.
\label{param:table}}
\end{table}

\begin{table}
\begin{center}
\begin{tabular}{cccc}
DJT energy gain 	&	this work	&
Ref.~\protect\cite{ob72,ob96}	&	exact \protect\cite{Reno}\\
\hline
order $k_{\rm eff}^2$	&	$-\frac 12 \sum k_i^2 \omega_i=-682.4$	&	the same	&	\\
order $k_{\rm eff}^0$	&	$\sum (\nu_j -\omega_j)=-722.8$	& $
-\frac 12 \frac{{\overline {\omega^2}}+{\omega_{\rm eff}}^2}{\omega_{\rm eff}}
=-713.4$& \\
pseudorotation $k_{\rm eff}^{-2}$ term 	&
$\left(3 \sum k_i^2 \omega_i^{-1}\right)^{-1}=60.1$	&	the same&	\\
Total gain (no pseudorotation)	&	-1405.1	&	-1395.8	&	\\
Total gain (including pseudorotation)&	-1345.0	&	-1335.7	&    -1125.7\\
\end{tabular}
\end{center}
\caption{The relevant ground-state energetics (in cm$^{-1}$) for the
many-modes DJT of C$_{60}^-$, in the present model and as obtained in the
cited works by  O'Brien.  The classical potential-lowering term, the
lowest order quantum corrections, and the pseudorotational contribution are
listed.  Finally, the strong-coupling energy gains (up to order $k_{\rm
eff}^{0}$, and including the pseudorotational correction) are
compared to the result of Lanczos diagonalization.
\label{GS:table}}
\end{table}

\noindent

\noindent

\noindent

\noindent

%\bibliographystyle{unsrt}
%\bibliography{biblio}

\begin{thebibliography}{10}
\bibitem{pseudosum:note}
{ This statement is equivalent to the nontrivial relation $\sum_{i>1} c_i^2 /
  \omega_i' = \omega_{\rm eff} - k_{\rm eff}^2/\sum k_i^2/\omega_i$, where the
  coefficients $c_i$ and frequencies $\omega_i'$ are those defined in
  Ref.~\cite{ob72}. This formula permits to write explicitly all the
  coefficients in the expansion Eq.~(44) of Ref.~\cite{ob72} in terms of
  momenta $\overline {\omega^m}$ ($m=-1,0,1,2$) of the distribution of
  frequencies}.
\end{thebibliography}

\end{document}